# Superconductivity and quantum oscillations in crystalline Bi nanowires

Mingliang Tian*, Jian Wang*, Qi Zhang, Nitesh Kumar, Thomas E. Malouk and Moses H. W. Chan*

*Center for Nanoscale Science, The Pennsylvania State University, University Park, Pennsylvania 16802-6300.*

*Emails: tian@phys.psu.edu (tian); juw17@psu.edu (wang); chan@phys.psu.edu(chan)*

While bulk bismuth (Bi) is a semimetal, we have found clear evidence of superconductivity in a crystalline 72 nm diameter Bi nanowire below 1.3 K. In a parallel magnetic field (H), the residual resistance of the nanowire below $T_c$ displays periodic oscillations with H, and the period corresponds to the superconducting flux quantum. This result provides evidence that the superconductivity comes from the interface shell between Bi and the surface oxide. In a perpendicular H, the resistance in the superconducting state shows Shubnikov-de Haas (SdH) oscillations, a signature of a normal metal. These results indicate a novel coexistence of Bosonic and Fermionic states in the surface shell of nanowires.

Extremely low carrier density, small effective carrier mass and very long electron mean free path [1] distinguish Bi from other metals as particularly suitable for studying quantum phenomena [2,3]. While bulk Bi is a semimetal down to at least 50 mK [4], its electronic properties in a confined geometry are more complex. Quench-condensed granular Bi films were found to be superconducting with a grain size dependent $T_c$ ranging from 2.0 K-5.5 K [5, 6]. Interestingly, granular Bi nanowires with a grain size of 10-15 nm fabricated by electro-deposition into porous membranes show a much higher $T_c$ of 7.2 K or 8.3 K [7]. These $T_c$'s coincide with those of high-pressure phases, Bi-III and Bi-V [8]. In contrast to granular systems, single crystalline Bi nanowire was predicted [9] to be a semiconductor when its diameter is below a critical value. However, extensive experimental studies [10-12] failed to confirm this prediction. There is evidence that the transport behavior of Bi nanowires is dominated by surface effects. In 55 nm and 75 nm single-crystalline Bi nanowire arrays, Nikolaeva *et al.* [12] observed Aharonov-Bohm (AB)-type [13] magnetoresistance oscillations due to quantum interference of phase coherent electrons along the bifurcating trajectories on the highly conducting surface of the wires. In an array of 30 nm diameter nanowires, Huber *et al*. [11] found isotropic Shubnikov-de Haas (SdH) oscillations, in contrast to the ordinarily anisotropic SdH oscillations in bulk Bi. This led the authors to propose a scenario in which the 30 nm Bi nanowire is metallic in its entire volume with a spherical Fermi surface (FS) due to surface states, which inhibits a semimetal to semiconductor transition.

This work explores the transport properties of individual single-crystalline Bi nanowires of 72 nm diameter. Remarkably, we found the nanowire displays superconductivity below 1.3 K. Under a parallel magnetic field, the residual resistance of the wires shows periodic oscillations with H deep in the superconducting state. The period of the oscillations corresponds to the superconducting flux quantum, $\Phi_0$, reminiscent of Little-Parks (LP) oscillations in a "hollow" superconducting cylinder [14]. The LP-like oscillations unambiguously suggest that the observed superconductivity in single-crystalline Bi wire originates from a surface shell of the Bi nanowire. Little-Parks oscillations were observed previously in hollow AuIn and Al cylinders with diameter as small as 150 nm [15]. In this experiment we show the phenomenon is realized in a shell of only 72 nm. Interestingly, when the field is aligned perpendicular to the wire axis, the residual resistance showed clear oscillations with 1/H periodicity below 0.6 K. The 1/H periodicity is indicative of SdH effect due to Landau quantization of the orbiting quasi-particle energy of a metal. These data suggest the coexistence of superconducting and metallic states in the surface shell near the oxidation layer.

The nanowires were fabricated by electrochemically depositing Bi into custom-made porous anodic aluminum oxide membranes [16]. The electrolyte was prepared as: 10g/L $BiCl_3$+50g/L NaCl +50g/L tartaric acid +100g/L glycerol. The milky aqueous solution was then made clear by adjusting the pH to 1-2 with concentrated hydrochloric acid (HCl). The electrodeposition was carried out at a constant potential of - 0.18 V, relative to an Ag/AgCl reference electrode at room temperature. A pure Pt wire was used as the positive counter electrode and an Ag film, evaporated onto one side of the membrane prior to the deposition was used as the working electrode. Free standing nanowires were obtained by dissolving the membrane with 2 M aqueous NaOH solution, followed by precipitating the wires via centrifugation. X-ray diffraction (XRD) studies were carried out on arrays of wires still embedded inside the porous membrane and transmission electron microscopy (TEM) were carried out on wires released from the membranes. In our TEM studies, we usually start with a quick check on the uniformity and morphology of typically 100 wires dispersed on the TEM grid. After the quick examination, we typically study 15-20 wires systematically at different magnifications before picking 3-4 wires for careful electron diffraction studies including the recording of high resolution images. Figure 1(a) shows an enlarged TEM image of a randomly selected Bi nanowire. The nanowire has a diameter of 79 ± nm and displays a bamboo-like morphology along its length. A high resolution TEM image (Fig. 1(b)) taken near the "bamboo" stripe indicated that the stripes shown in Fig. 1(a) are actually small angle twinning boundaries with a twinning plane of (10$\underline{1}$4). This result was confirmed by XRD on Bi wire arrays, where the (104) is the dominant peak in the spectra. An oxide layer of approximately ~ 3.7 ± 0.5 nm, probably formed after the nanowires were released from the membrane, is visible on the surface. Therefore, the actual inner diameter (d) of the Bi nanowire inside the oxide shell is about 72.0 ± 1.0 nm. No granular wires were ever seen.

Standard 4-probe transport measurements were carried out on four similarly fabricated individual Bi nanowires harvested from four different membranes (samples N1, N2, N3 and N4) in a physical property measurement system (PPMS) cryostat equipped with a dilution refrigerator insert and a 9 T superconducting magnet. Four Pt strips of 100 (width)×100 (thickness) nm $^2$ were deposited onto the Bi nanowires as electrodes using the focused-ion beam (FIB) technique (see inset of Fig. 1(c). It was found that the insulating Bi-O oxide layer covering the Bi wire is reliably penetrated by the deposition of Pt electrodes and an ohmic contact can be obtained between the Pt lead and the Bi wire. This was confirmed by linear V-I characteristics in the normal state of the wire. The maximum spreading distance of Pt along the nanowires was found to be less than 300 nm beyond the intended position through a profile analysis of the TEM energy dispersive x ray study. For all four samples, the distance (L) between the inner edges of the two voltage electrodes was kept larger than 2.0 µm.

Resistance (R) vs. temperature (T) curves of sample N1 (L = 2.35 µm) and N4 (L = 4.9 µm) are shown in Fig. 1(c) in log-log scale. Just below 1.3 K, the resistance of the nanowire shows a sharp drop, decreasing by three orders of magnitude and eventually reaching a value beyond the resolution of our system below 0.3 K. The log R-log T plot shows a shoulder around 0.67 K. Very similar behavior was found in sample N4 and its resistivity is consistent with that of N1 to within 5%. By applying a perpendicular H, the transition temperature of both samples was pushed to low temperatures and finally below 50 mK at 40 kOe, as shown in Figs.1(d) and 1(e), respectively. The three orders of magnitude drop in R down to zero is a clear signature of a superconducting transition at 1.3 K in both N1 and N4. The shoulder near 0.67 K is likely related to the superconducting phase at 0.64 K reported by Ye *et al.* [17] for single-crystal Bi nanowires still embedded in a porous membrane. In the experiment of Ye *et al.*, the total resistance drop from 0.64 K down to 50 mK was found to be 30% of its normal state value. To our knowledge, this is the only other experimental evidence showing superconductivity in small diameter single-crystalline Bi nanowires.

Figure 2 shows voltage-current (V-I) characteristics of sample N1, measured at different temperatures. At 50 mK, a sharp discontinuous jump in voltage was seen at 0.14 µA (0.12 µA in N4). This value corresponds to a superconducting critical current density of $j_s = 3.4 \times 10^3$ A/cm$^2$ if we assume that the current passes uniformly through the entire cross-section area of the wire. This value is 2 to 3 orders of magnitude smaller than that observed in typical superconducting nanowires ($j_s \sim 10^5$-$10^6$ A/cm$^2$ in Sn, Nb and Al nanowires) at nearly the same temperature [18-20]. The critical current decreases with increasing temperatures and the crossover from superconducting to normal state becomes more rounded.

R vs. H isotherms measured with H aligned perpendicular and parallel to the wire axis are shown in Figs. 3 (a) and 3 (b), respectively. The nanowire exhibits strong anisotropic magnetic behavior with a broad phase transition from the superconducting to normal state, where the magnitude of the perpendicular critical field at 0.1 K, $H_{ct}$ ~ 34 kOe, defined from the point at which the resistance of the nanowire is restored to 97.5 % of its normal state value, is approximately one half that of the parallel critical field, $H_{cp}$ ~ 67 kOe. This demagnetization effect is consistent with the expectations for a thin superconducting cylinder [21].

The most remarkable feature of these two figures is the series of resistance oscillations observed at magnetic fields and temperatures well below their critical values. To show clearly the resistance oscillations, a smooth R vs. H background was subtracted from each of the R-H isotherms. The subtracted curves $\Delta R_{//}$ at different temperatures show oscillations periodic with H (Fig.3 d). The vertical dashed lines, separated by $\Delta H$ in multiples of 5.85 kOe from H = 0 Oe are found to locate the majority of the local minima of the $\Delta R_{//}$ - H isotherms. At higher temperatures, oscillations in R are extended to lower H but become difficult to resolve at higher fields. For T < 0.4 K, additional peaks appear at fields above 40 kOe.

In a superconducting system, the periodic oscillations of resistance as a function of parallel external magnetic field are expected for a superconducting ring or a hollow cylinder. This phenomenon, named the Little-Parks (LP) effect [14] is a consequence of "fluxoid" quantization, where the oscillations in magnetic fields are characterized by the integer multiples of the superconducting flux quantum, i.e., $H(\pi d^2/4) = n\Phi_0$, where $\Phi_0 = 2.07 \times 10^7$ G.cm$^2$, d is the diameter of the cylinder and n is an integer. In this model, $\Delta H$ = 5.85 kOe results in d = 67.0 nm. This value is close to 72 nm, the wire diameter as determined by TEM. Since d corresponds to the inner diameter of the "hollow" cylinder, comparison to the physical diameter results in a model of the Bi nanowire as a superconducting shell with a thickness of ~2.5 nm and a non-superconducting core of ~ 67.0 nm. The fact that the minima in $\Delta R_{//}$ coincide with integer multiples of the superconducting flux quantum provides strong indication that the superconductivity in these crystalline Bi nanowires comes from a thin cylindrical shell between the insulating surface oxide layer and the crystalline inner core of metallic Bi. This thin shell model is consistent with the observation of unusually small critical current density in Bi nanowires (Fig. 2), relative to the value expected if the entire wire was superconducting. Transport measurements were also made on an ensemble of crystalline Bi nanowire still imbedded inside the membrane. No evidence of superconductivity was found down to 0.47 K [7]. Since Bi wires inside the membrane have no oxide layer, this result supports the notion that the oxide layer is responsible in inducing superconductivity in the wire.

A significant deviation from standard LP behavior in our results is that the oscillations are observed deep in the superconducting state, far below $T_c$, whereas the conventional LP effect is expected only at the temperatures very close to its $T_c$. Also, it requires the superconducting phase coherence length, $\xi(0)$, to be much smaller than the diameter, d, of the cylinder and the penetration depth, $\lambda_p(0)$, to be larger than the wall thickness, t, of the cylinder. According to the G-L theory, the superconducting phase coherence length in the cylindrical shell of Bi wire was estimated to $\xi(0)$ ~ 67.5 nm, based on the formula, $\xi(0) = \sqrt{3}\Phi_0 / \pi t H_c(0)$ [21], with $H_c(0)$ ~ 67 kOe and t ~ 2.5 nm. This value is on the order of the diameter, d ~ 67 nm of the inner core of the cylinder. In an ultrathin cylinder of d < $\xi(0)$, Liu et al. [15] reported the existence of destructive regimes at the odd half integer multiples of $\Phi_0$, characterized by the loss of the global phase coherence. The resistance peaks in these ultrathin cylinders could be seen even at zero temperature, in contrast to those in a large superconducting ring or cylinder. Our system is probably in the cross-over regime between the conventional LP and that reported by Liu et al..

Oscillations in resistance are also seen in a perpendicular field (Fig. 3a). However, the period of these oscillations is not periodic with H (Fig. 3c). Fig.4 shows $\Delta R_\perp$ -1/H curves and the inset is the blow-up of the plot in the field region H >7.5 kOe. It appears that the R-oscillations are quasi-periodic with 1/H between 7.5 kOe and 25 kOe. The dashed lines pinpoint all the local minima of the oscillations in $\Delta R_\perp$ except for the minima near 1/H= 0.115 kOe$^{-1}$ for temperatures below 0.4 K. Oscillations in R that are periodic with 1/H are characteristic signatures of SdH oscillations. We note that 1/H oscillations were also found in N4 but with different periodicity. As a result, the observation of SdH oscillations is usually taken as clear-cut evidence for the existence of the FS, and is unexpected for a system in the superconducting state.

It is tempting to ascribe the SdH effect not to the superconducting interface shell but to the inner non-superconducting core. This speculation is however inconsistent with our experimental data. First, the parallel contribution of the normal inner core to the total conductance is expected to be negligibly small compared to that of the nearly fully superconducting shell. Thus, any variations or oscillations in the resistance of the inner core would be very difficult to detect far below $H_c$ (or $T_c$). Second, the SdH signatures were seen only deep in the superconducting state for T<0.7 $T_c$. If the 1/H oscillations originate from the normal core, they should not be sensitive to the onset of superconductivity and should be observed both above and below $T_c$. Hence, the most reasonable explanation is that the SdH oscillations also have their origin from the same interface shell, and there is a novel coexistence of superconducting and metallic phases.

SdH oscillations were reported in some of highly anisotropic type-II superconductors in the reversible vortex states [22], where vortices are free to move and a finite resistance exists. A central feature in these systems is that the SdH was seen in both the superconducting and normal states with the same periodicity. The amplitude of oscillations in the superconducting phase showed a significant attenuation due to the opening of a superconducting gap. In our experiments, SdH oscillations were found only deep in the superconducting phase far below $H_c$.

We note that samples N2 (L = 3.0 μm) and N3 (L = 2.1μ m) also showed a superconducting drop in resistance, but the onset of the transition was found near 1.9 ± 0.2 K and the resistance drop ended at 75% and 12% of its normal state value, respectively. Although the resistance oscillations below $T_c$ were seen in N2 and N3, no LP-like or SdH-like oscillations can be resolved or identified. LP oscillations require the formation of an unbroken superconducting cylindrical shell around the wire. This is evidently not satisfied in N2 and N3 where the resistance drops incompletely. We noted that SdH oscillations were not seen below 7.5 kOe. This is because the low field oscillations correspond to high Landau-level quantization, $\nu = (2\pi l_B / \lambda_F)^2$ [23], with $l_B = (\hbar/eH)^{1/2}$ as the magnetic length and $\lambda_F$, the Fermi wave length. Such a quantization is inhibited when $l_B$ exceeds the radius of the wire. For a wire with diameter of 72 nm, the minimum field for SdH corresponds to 5.0 kOe, not inconsistent with our observation of 7.5 kOe. The number of the oscillations depends on the extremal cross-sectional area, $A_f$ of the Fermi surface (FS) normal to H. It requires specific alignment of the anisotropic FS with respect to H. It is perhaps a fortunate coincidence that in sample N1 the magnetic field is indeed aligned along a specific preferred direction of the FS. We should note that we do not fully understand why the best periodicity is found near 0.4 K and 0.5 K. It is possible that these less perfect correlation of the oscillations in N1-N4 at the magnetic fields are a consequence of the variations of the FS in the normal regions with respect to applied H and the inhomogeneity of the superconductivity in the interface shell.

In summary, we report the observation of superconductivity in a crystalline Bi nanowire of 72 nm in diameter below 1.3 K. Both Little-Parks-like oscillations in parallel fields and SdH oscillations in perpendicular H were found. These data suggest a completely unexpected and novel coexistence of Bosonic (i.e., superconducting) and Fermionic (i.e., metallic) states in the surface shell of the Bi nanowire below $T_c$.

This work was supported by the Center for Nanoscale Science (Penn State MRSEC) funded by NSF under Grant No. DMR-0820404.

**Figure captions**

Figure 1. (a) TEM image of a randomly selected Bi nanowire showing bamboo-like morphology highlighted by arrows. (b) High-resolution TEM image showing a twinning boundary.(c) R-T plot at H=0 Oe for a 72 nm diameter Bi nanowire (samples N1 and N4) on a log-log scale.The inset is an SEM image of Pt electrodes on the Bi nanowire. (d) and (e) are respectively the R-T curves of samples N1 and N4 at different magnetic fields applied perpendicular to wire axis.

Figure 2. V-I curves for sample N1 at different temperatures. Inset shows $I_c$- T dependence.

Figure 3. (a) and (b) show the R-H curves for sample N1 at different temperatures for magnetic fields applied perpendicular and parallel to the wire axis, respectively. (c) and (d), are respectively, the curves obtained by subtracting smooth background from (a) and (b). Dashed lines in (d) indicate the positions of fluxoid quantization as predicted by $H(\pi d^2/4) = n \Phi_0$.

Figure 4. $\Delta R_\perp$ - 1/H curves deduced from the data in Fig.3 (c). The inset shows the blow-up of the plot in the field regime, H >7.5 kOe. Quasi-periodic oscillations of R with 1/H are clearly seen, especially for the curves at 0.4 K and 0.5 K. Dashed lines, separated by $\Delta(1/H)=0.0176$ kOe$^{-1}$ show good correlation with the minima of resistance oscillations as predicted by SdH effect.

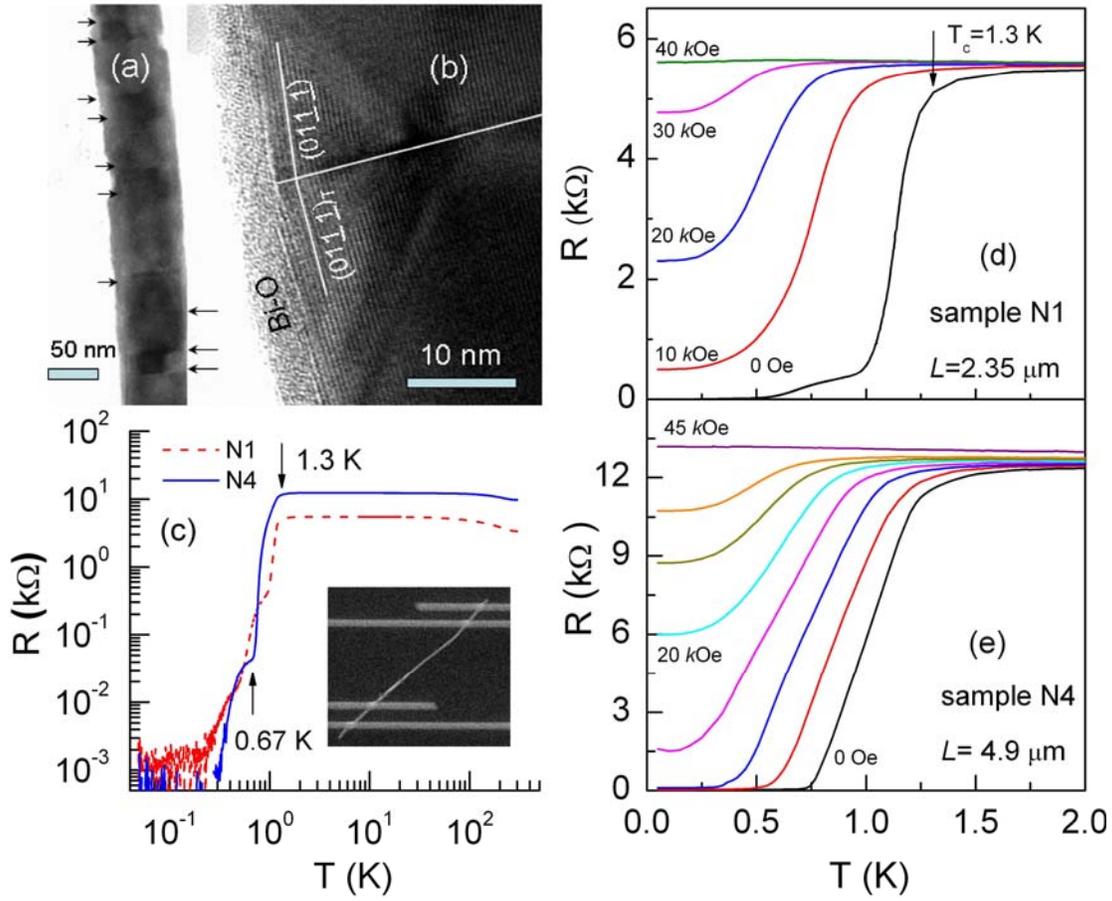

Figure 1. (a) TEM image of a randomly selected Bi nanowire showing bamboo-like morphology highlighted by arrows. (b) High-resolution TEM image showing a twinning boundary. (c) R-T plot at H=0 Oe for a 72 nm diameter Bi nanowire (samples N1 and N4) on a log-log scale. The inset is an SEM image of Pt electrodes on the Bi nanowire. (d) and (e) are respectively the R-T curves of samples N1 and N4 at different magnetic fields applied perpendicular to wire axis.

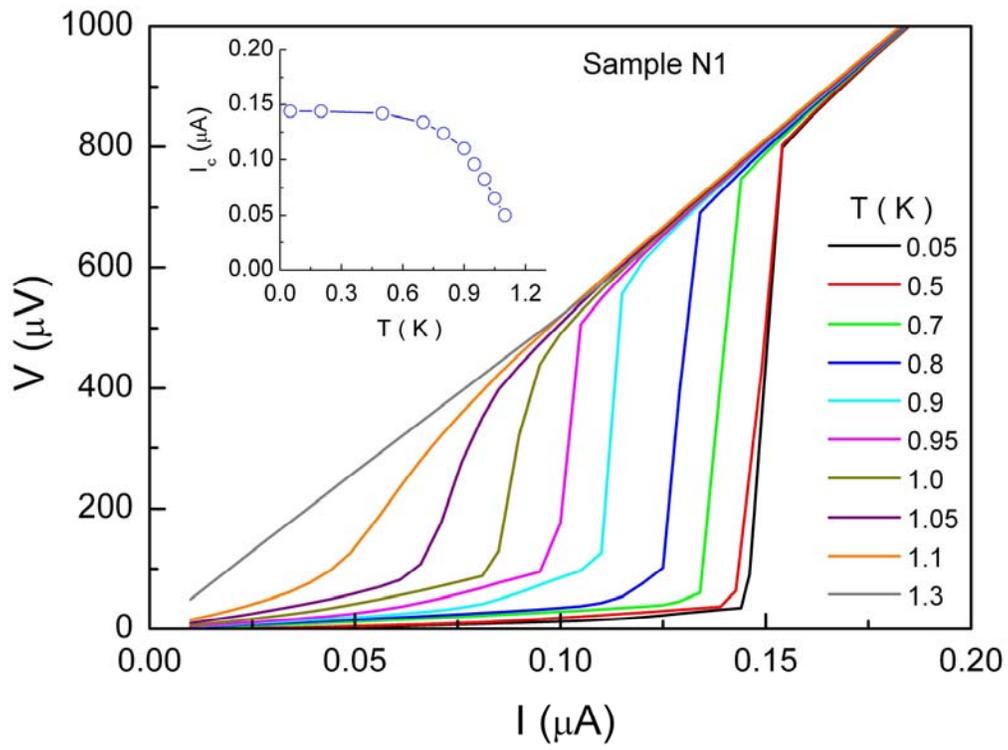

Figure 2. V-I curves for sample N1 at different temperatures. Inset shows $I_c$- T dependence.

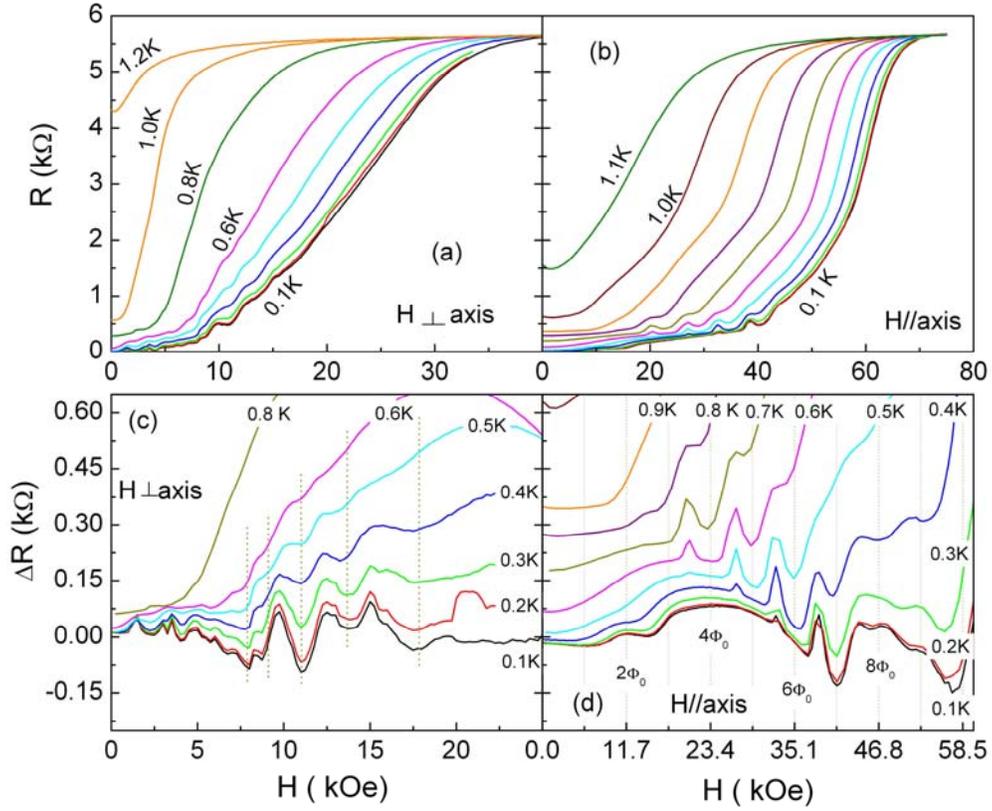

Figure 3. (a) and (b) show the R-H curves for sample N1 at different temperatures for magnetic fields applied perpendicular and parallel to the wire axis, respectively. (c) and (d), are respectively, the curves obtained by subtracting smooth background from (a) and (b). Dashed lines in (d) indicate the positions of fluxoid quantization as predicted by $H(\pi d^2/4) = n\, \Phi_0$.

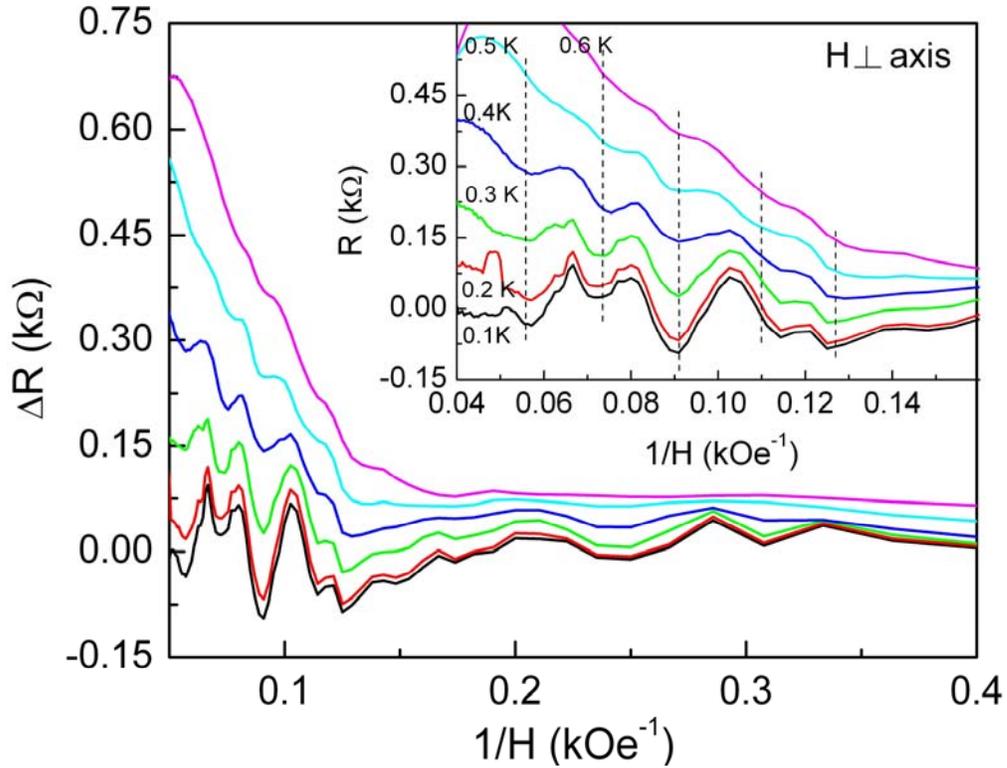

Figure 4. ΔR$_\perp$ - 1/H curves deduced from the data in Fig.3 (c). The inset shows the blow-up of the plot in the field regime, H >7.5 kOe. Quasi-periodic oscillations of R with 1/H are clearly seen, especially for the curves at 0.4 K and 0.5 K. Dashed lines, separated by Δ(1/H)=0.0176 kOe$^{-1}$ show good correlation with the minima of resistance oscillations as predicted by SdH effect.